\documentclass[12pt]{article}
\usepackage{amsmath}
\usepackage{bm}
\usepackage{color}

\begin{document}
\begin{center}
\begin{large}
{\bf Features of description of composite system's motion in twist-deformed space-time}\\
\end{large}
\end{center}

\centerline {Kh. P. Gnatenko \footnote{E-Mail address: khrystyna.gnatenko@gmail.com}}
\medskip
\centerline {\small \it Ivan Franko National University of Lviv, Department for Theoretical Physics,}
\centerline {\small \it 12 Drahomanov St., Lviv, 79005, Ukraine}
\centerline {\small \it  Laboratory for Statistical Physics of Complex Systems}
\centerline {\small \it  Institute for Condensed Matter Physics, NAS of Ukraine, Lviv, 79011, Ukraine}

\abstract{  Composite system made of $N$ particles is considered in twist-deformed space-time. It is shown that in the space the motion of the center-of-mass of the system depends on the relative motion. Influence of deformation on the motion of the center-of-mass of composite system is less than on the motion of individual particles and depends on the system's composition. We conclude that if we consider commutation relations for coordinates of a particle to be proportional inversely to its mass, the commutation relations for coordinates of composite system do not depend on its composition  and are proportional inversely to system's total mass, besides the motion of the center-of-mass is independent of the relative motion. In addition we find that inverse proportionality of parameters of noncommutativity to mass is important for considering coordinates in twist-deformed space as kinematic variables and for preserving of the weak equivalence principle.

Key words: twist-deformed space-time, composite system, weak equivalence principle, kinematic variables.
PACS numbers: 11.10 Nx, 11.90.+t
}

\section{Introduction}

In recent years studies of idea that space coordinates might be noncommutative have attracted much attention owing to development of String Theory and Quantum Gravity \cite{Witten,Doplicher}.

 Much attention has been devoted to studies of quantum and classical physical problems (harmonic oscillator \cite{Hatzinikitas,Kijanka,Jing,Smailagic,Smailagic1,Djemai,Giri,Geloun,Nath,GnatenkoJPS17},  particle in gravitational field \cite{Bertolami1,Bastos},  composite systems \cite{Ho,Bellucci,GnatenkoPLA13,Djemai,GnatenkoPLA17,Daszkiewicz},  quantum fields  \cite{Balachandran10,Balachandran11} and many others) in a space with canonical noncommutativity of coordinates
\begin{eqnarray}
[X_{\mu},X_{\nu}]=i\hbar\theta_{\mu\nu},\label{can}
\end{eqnarray}
with $\theta_{\mu\nu}$ being constants.

In papers
\cite{DaszkiewiczMPLA12,DaszkiewiczPS} it is shown that in the case of Newton-Hooke Hopf algebras
 the noncommutativity of the form
\begin{eqnarray}
[t,X_i]=0,\\{}
[X_i,X_j]=i\hbar f_{\pm}\left(\frac{t}{\tau}\right)\theta_{ij},{}\label{nt}{}
\end{eqnarray}
is provided by the twist deformation. Here  $\tau$ is a time-scale parameter, functions $f_{+}\left(\frac{t}{\tau}\right)=f\left(\sinh\left(\frac{t}{\tau}\right),\cosh\left(\frac{t}{\tau}\right)\right)$, $f_{-}\left(\frac{t}{\tau}\right)=f\left(\sin\left(\frac{t}{\tau}\right),\cos\left(\frac{t}{\tau}\right)\right)$  correspond to N-enlarged Newton-Hooke Hopf algebras \cite{DaszkiewiczPS}. Parameters $\theta_{ij}$ are considered to be constants or $\theta_{ij}=\theta^k_{ij}X_k$  with $\theta^k_{ij}$ being constants, $i,j,k=(1,2,3)$, $\tau$ is a time scale parameter.  Constant $\hbar$ in (\ref{nt}) is written for convenience.  In the limit $\tau\rightarrow\infty$ form  (\ref{nt}) commutation relations  of canonical type (\ref{can}), Lie type
$[X_{\mu},X_{\nu}]=i\hbar\theta^{\rho}_{\mu\nu}X_{\rho},$ \cite{Lukierski18,Meljanac,Miao} and commutation relations with quadratic noncommutativity
$[X_{\mu},X_{\nu}]=i\theta^{\rho\tau}_{\mu\nu}X_{\rho}X_{\tau},$ \cite{Tureanu,Lukierski06} can be reproduced  \cite{DaszkiewiczMPLA09}  (here
 $\theta^{\rho}_{\mu\nu}$, $\theta^{\rho\tau}_{\mu\nu}$ are constants).

The results of studies of  many-particle problems in the frame of noncommutative algebra open possibility to investigate effect of space quantization on the properties of wide class of physical systems. In recent paper \cite{DaszkiewiczPS} a two-particle system in Coulomb potential was studied in twist-deformed space  and a helium atom as an example of the two-particle system was considered.
In \cite{Daszkiewicz1} many-body quantum mechanics for twisted N-enlarged
Newton-Hooke space-times was studied. The author provided Schroedinger equation for arbitrary
stationary potential. As an example a system of particles in Coulomb field  interacting by Coulomb potential was considered.

In the present paper we examine features of description of motion of the center-of-mass of composite system and relative motion in twist-deformed space-time. We show that influence of the deformation on the motion of the center-of-mass depends on the system's mass and  composition. We also conclude that the motion of the center-of-mass is not independent of the relative motion in twist-deformed space.  We find that relation of parameters of twist-deformed algebra with mass opens possibility to preserve independence of motion of the center-of-mass of the relative motion, independence of motion of composite system (particle) on its mass and composition in gravitational field (recovering of the weak equivalence principle), independence of coordinates on mass in twist-deformed space (coordinates can be considered as kinematic variables).

The article is organized as follows. In the Section 2 features of twist-deformed algebra for coordinates and momenta of the center-of-mass and the relative motion are discussed. Section 3 is devoted to studies  of motion of a particle (composite system) in gravitational field  and  examining of implementation of the weak equivalence principle in twist-deformed space. Representation for coordinates satisfying twist-deformed algebra is considered in Section 4 and the problem of kinematic variables is discussed. Conclusions are presented in Section 5.

\section{Twist-deformed algebra for coordinates and momenta of the center-of-mass and the relative motion }

Let us consider a many-particle system composed by $N$ particles of masses $m_a$ and described by the following Hamiltonian
\begin{eqnarray}
 H_s=\sum_a\frac{( {\bf P}^{(a)})^{2}}{2m_a}+\frac{1}{2}\mathop{\sum_{a,b}}\limits_{a\neq b}V_{int}(|{\bf X}^{(a)}-{\bf X}^{(b)}|),\label{form777}
\end{eqnarray}
in twist-deformed space-time. Here indexes $a$ and $b$ label the particles, coordinates $X^{(a)}_i$, and momenta $P^{(a)}_i$ satisfy twist-deformed commutation relations
\begin{eqnarray}
[X^{(a)}_i,X^{(b)}_j]=i\hbar\delta_{ab}f\left(\frac{t}{\tau}\right)\theta^{(a)}_{ij},{}\label{ntp}\\{}
[t,X^{(a)}_j]=0, \ \ [P^{(a)}_i,P^{(b)}_j]=0,\label{nt2}\\{}
[X^{(a)}_i,P^{(b)}_j]=\delta_{ij}\delta_{ab}i\hbar.\label{nt1}
 \end{eqnarray}
here $i,j=(1,2,3)$, $\theta^{(a)}_{ij}$ are parameters of noncommutativity corresponding to particle with mass $m_a$, which are considered to be constants, and notation $f\left(\frac{t}{\tau}\right)=f_{\pm}\left(\frac{t}{\tau}\right)$ is used. Note, that we study a general case when parameters of noncommutativity are supposed to be different for different particles.

Total momentum defined as
\begin{eqnarray}
 \tilde{{\bf P}}=\sum_{a}{\bf P}^{(a)}\label{m}
\end{eqnarray}
is an integral of motion. It is easy to check that
\begin{eqnarray}
[\tilde{{P}}_i,H_s]=0.\label{int}
\end{eqnarray}
Therefore, one can introduce coordinates of the center-of-mass in the traditional way
\begin{eqnarray}
\tilde{{\bf X}}=\sum_{a}\mu_{a}{\bf X}^{(a)}\label{m1}
\end{eqnarray}
here $\mu_a=m_{a}/M$, $M=\sum_{a}m_a$. Coordinates and momenta in (\ref{m}), (\ref{m1}) satisfy commutation relations (\ref{nt})-(\ref{nt1}). Upon straightforward calculations one can write the following commutators for coordinates of the center-of-mass and total momenta
\begin{eqnarray}
[\tilde{X}_i,\tilde{X}_j]=i\hbar f\left(\frac{t}{\tau}\right)\sum_{a}\mu_a^2\theta^{(a)}_{ij},\label{cm}{}\\{}
[\tilde{P}_i,\tilde{P}_j]=0,\\{}
[\tilde{X},\tilde{P}_j]=i\hbar\delta_{ij}.\label{cm1}
\end{eqnarray}
Note that in comparison with (\ref{ntp}) commutator for coordinates of the center-of-mass (\ref{cm}) depends on masses of particles which form the system, therefore it depends on the system's composition. It is worth noting that for the case of a system made of $N$ particles of masses $m$ and parameters $\theta_{ij}$ one obtains
\begin{eqnarray}
[\tilde{X}_i,\tilde{X}_j]=\frac{i\hbar}{N}f\left(\frac{t}{\tau}\right)\theta_{ij}.\label{red}
\end{eqnarray}
So, from (\ref{red}) we have that relations of noncommutative algebra for coordinates of the center-of-mass contain parameters of noncommutativity which are $N$ times smaller than parameters of noncommutativity corresponding to individual particles. From this follows that influence of features of space structure on  macroscopic systems is less than on  elementary particles.

 Introducing  momenta and coordinates of the relative motion in traditional way
 \begin{eqnarray}
\Delta{\bf P}^{{a}}={\bf P}^{(a)}-\mu_{a}\tilde{{\bf P}},\ \ {\Delta\bf X}^{(a)}={\bf X}^{(a)}-\tilde{{\bf X}},\label{06}
\end{eqnarray}
on the basis of relations (\ref{nt})-(\ref{nt1}) one can calculate
\begin{eqnarray}
[\Delta{X}_i^{(a)},\Delta{X}_j^{(b)}]= i\hbar f\left(\frac{t}{\tau}\right)\left(\delta_{ab}\theta^{(a)}_{ij}-\mu_{a}\theta^{(a)}_{ij}-\mu_{b}\theta^{(b)}_{ij}+\sum_{d}{\mu_{d}^{2}}\theta^{(d)}_{ij}\right),\\{}
[\Delta{X}^{(a)}_i,\Delta{P}^{(b)}_j]=i\hbar(\delta_{ab}-\mu_b),\\{}
[\Delta{P}_i^{(a)},\Delta{P}_i^{(b)}]=0.
\end{eqnarray}
It is important to note that because of noncommutativity (\ref{ntp}) one has that commutators for coordinates of the center-of-mass and coordinates of the relative motion do not vanish. We can calculate that the commutation relations read
\begin{eqnarray}
[\Delta{X}_i^{(a)},\tilde{X}_j]= i\hbar f\left(\frac{t}{\tau}\right) \left({\mu_{a}}\theta^{(a)}_{ij}-\sum_{d}\mu_{d}^{2} \theta^{(d)}_{ij}\right).\label{cr}
\end{eqnarray}
For relative and center-of-mass momenta one has $[\Delta{P}_i^{(a)},\Delta{P}_i^{(b)}]=[\tilde{P}_i,\Delta{P}_j^{(b)}]=0.$ From (\ref{cr}) follows that one cannot consider motion of the center-of-mass of a system and relative motion as independent. Relative motion effects on the motion of the center-of-mass and vice versa.

At the end of this section we would like to stress that from (\ref{cm}) follows that for different composite systems (different composite particles, bodies) one has to consider parameters of noncommutativity to be dependent on their masses and compositions. It is worth mention that if we assume parameters of noncommutativity corresponding to a particle to be proportional inversely to its mass. Namely, if we consider the product $m_a\theta^a_{ij}$ to be a constant $\gamma_{ij}$ which is the same for different particles
\begin{eqnarray}
{m_a}\theta^{(a)}_{ij}=\gamma_{ij}=const,\label{cond}
\end{eqnarray}
one obtains that
\begin{eqnarray}
[\Delta{X}_i^{(a)},\tilde{X}_j]=0.\label{crz}
\end{eqnarray}
So, the motion of the center-of-mass is independent on the relative motion. In addition, if (\ref{cond}) holds, one has
\begin{eqnarray}
[\tilde{X}_i,\tilde{X}_j]=i\hbar f\left(\frac{t}{\tau}\right)\frac{\gamma}{M}.
\end{eqnarray}
So, the commutator for coordinates of the center-of-mass of a system does not depend on its composition and is proportional inversely to its total mass $M$.
Note that condition  (\ref{cond}) can be written also for effective parameter of noncommutativity
\begin{eqnarray}
\theta^{eff}_{ij}=\sum_{a}\mu_a^2\theta^{(a)}_{ij}\label{eff}
 \end{eqnarray}
 which describes the motion of the center-of-mass (\ref{cm}). In the case when parameters of noncommutativity of  particles are proportional inversely to their masses (\ref{cond}) the effective parameter of noncommutativity corresponding to a composite system is proportional inversely to  total mass of the system $M$, we have $\theta^{eff}_{ij}=\gamma_{ij}/M$.

 In the next section we will show that condition (\ref{cond}) is important for preserving of the weak equivalence principle in twist-deformed space-time.

\section{Implementation of the weak equivalence principle in twist-deformed space-time}

Implementation of equivalence principle was examined in the case of algebra with noncommutativity of coordinates of canonical type \cite{GnatenkoPLA13,Saha,Saha1}, algebra with noncommutativity of coordinates  and noncommutativity of momenta of canonical type \cite{Bastos1,Bertolami2,GnatenkoPLA17}, in rotationally-invariant noncommutative phase space of canonical type \cite{GnatenkoEPL18}, in deformed space with minimal length \cite{Tkachuk}.
Let us consider the weak equivalence principle in the case of twist-deformed space.
 So, let us consider a motion of a particle of mass $m$ in gravitational filed $V=V({\bf X})$ in a space with commutation relations (\ref{ntp})-(\ref{nt1}). The Hamiltonian reads
\begin{eqnarray}
H=\frac{{\bf P}^2}{2m}+mV({\bf X}).\label{hamilt}
\end{eqnarray}
Taking into account (\ref{ntp})-(\ref{nt1}) one can write the following equations of motion
 \begin{eqnarray}
\dot{X}_i=\frac{P_i}{m}+ f\left(\frac{t}{\tau}\right)m\theta_{ij}\frac{\partial V}{\partial X_j},\label{motion}\\
\dot{P}_i=-m\frac{\partial V}{\partial X_i},
 \end{eqnarray}
From the first equation one has that because of deformation the velocity of a particle in gravitational field depends on its mass. According to the weak equivalence principle the motion of a particle in gravitational field does not depend on its mass and composition. From these statements follows that in twist-deformed space the weak equivalence is not preserved.

We would like to stress that if we consider the parameters of noncommutativity to satisfy  relation (\ref{cond}) the equations of motion read
 \begin{eqnarray}
\dot{X}_i={P^{\prime}_i}+ f\left(\frac{t}{\tau}\right)\gamma_{ij}\frac{\partial V}{\partial X_j},\label{motion1}\\
\dot{P}^{\prime}_i=-\frac{\partial V}{\partial X_i},\label{motion2}
 \end{eqnarray}
here $P^{\prime}_i$ denotes $P^{\prime}_i=P_i/m$. From equations (\ref{motion1}), (\ref{motion2}) we can conclude that their solutions   ${X}_i(t)$,  ${P}^{\prime}_i(t)$ do not depend on mass. Therefore the weak equivalence principle is preserved if condition (\ref{cond}) holds.
Note that this conclusion can be generalized to the case of motion of a composite system (body) of mass $M$ in gravitational filed. The Hamiltonian of the system reads
\begin{eqnarray}
H=\frac{{\tilde{\bf P}}^2}{2M}+MV({\tilde{\bf X}})+H_{rel}.\label{hambody}
\end{eqnarray}
Where ${\tilde{\bf X}}$, ${\tilde{\bf P}}$ are coordinates and momenta of the center-of-mass, satisfying (\ref{cm})- (\ref{cm1}), and  $H_{rel}$ denotes terms which correspond to the relative motion. If condition (\ref{cond}) holds one  has that coordinates of the center-of-mass commute with the coordinates of the relative motion (\ref{crz}), therefore $[H,H_{rel}]=0$ and we can consider the motion of the center-of-mass independently on the relative motion.
The equations of motion read
 \begin{eqnarray}
\dot{{\tilde X}}_i={{\tilde P}^{\prime}_i}+ f\left(\frac{t}{\tau}\right){\gamma_{ij}}\frac{\partial V({\tilde{\bf X}})}{\partial {\tilde X}_j},\label{mcfm}\\
\dot{{\tilde P}}^{\prime}_i=-\frac{\partial V ({\tilde{\bf X}})}{\partial {\tilde X}_i},\label{mcfm1}
 \end{eqnarray}
here ${\tilde P}^{\prime}_i={\tilde P}_i/M$. So, due to condition (\ref{cond}) the solutions of equations  (\ref{mcfm}), (\ref{mcfm1}) do not depend on the mass of a system (body) and on its composition. Therefore, the weak equivalence principle is recovered.

We would like to note here that if condition  (\ref{cond}) does not hold, we have that motion of composite system in gravitational field depends on its total mass. In addition  the motion of composite system depends on its composition because of dependence of effective parameters of noncommutativity (\ref{eff}) on masses of particles which form the system. This dependence is additional evidence of violation of the weak equivalence principle in twist-deformed space-time.

In the next section we will show that condition (\ref{cond}) gives a possibility to obtain another important result, namely to consider noncommutative coordinates as kinematic variables.

\section{Representation for coordinates satisfying twist-deformed commutation relations and kinematic variables}

Coordinates and momenta which satisfy commutation relations (\ref{ntp})-(\ref{nt1}) can be represented as
\begin{eqnarray}
X^{(a)}_i=x^{(a)}_i-\frac{1}{2}f\left(\frac{t}{\tau}\right)\theta^{(a)}_{ij}p^{(a)}_j,{}\label{repp}\\{}
P^{(a)}_i=p^{(a)}_i,
\end{eqnarray}
with  coordinates $x^{(a)}_i$ and momenta $p^{(a)}_i$ satisfying
$[x^{(a)}_i,x^{(b)}_j]=[p^{(a)}_i,p^{(b)}_j]=0$, $[x^{(a)}_i,p^{(b)}_j]=\delta_{ij}\delta_{ab}i\hbar$.
Note that representation for coordinates (\ref{repp}) contains momenta $p^{(a)}_j=P^{(a)}_j$ which depend on mass. So, coordinates $X^{(a)}_i$ cannot be considered as kinematic variables.

We would like to stress that if condition (\ref{cond}) holds one has
\begin{eqnarray}
X^{(a)}_i=x^{(a)}_i-f\left(\frac{t}{\tau}\right)\frac{\gamma_{ij}p^{(a)}_j}{2m_a}.{}\label{rep1}{}
\end{eqnarray}
So, the coordinates do not depend on mass and can be considered as kinematic variables due to  relation (\ref{cond}).

For coordinates and momenta of the center-of-mass taking into account their definitions (\ref{m}), (\ref{m1})  and using (\ref{repp}), one can write
\begin{eqnarray}
{\tilde X}_i=\sum_a\mu_aX^{(a)}_i= \tilde{x}_i-\frac{1}{2}f\left(\frac{t}{\tau}\right)\sum_a\mu_a\theta^{(a)}_{ij}p^{(a)}_j,\label{repcm}\\
\tilde{P}_i=\sum_a{P}^{(a)}_i=\tilde{p}_i,\label{repcm1}
\end{eqnarray}
Here we use notations $\tilde{x}_i=\sum_a\mu_a x^{(a)}_i$, $\tilde{p}_i=\sum_a{p}^{(a)}_i$. For coordinates  ${\tilde x}_i$, and momenta  ${\tilde p}_i$  the following relations hold $[{\tilde x}_i,{\tilde x}_j]=[{\tilde p}_i,{\tilde p}_j]=0$, $[{\tilde x}_i,{\tilde p}_j]=i\hbar\delta_{ij}$. It is easy to check that coordinates represented as (\ref{repcm}) satisfy (\ref{cm}).

Note that from (\ref{repcm}) (\ref{repcm1}) follows that coordinates of the center-of-mass  of a system depend on momenta of particles $p^{(a)}_j=P^{(a)}_j$ forming it and  therefore depend on the total momenta and momenta of the relative motion
\begin{eqnarray}
{\tilde X}_i=\tilde{x}_i-\frac{1}{2}f\left(\frac{t}{\tau}\right)\sum_a\mu_a\theta^{(a)}_{ij}(\Delta P^{(a)}_j+\mu_a{\tilde P}_j).
\end{eqnarray}
Note that the situation is changed if we consider the condition (\ref{cond}) to be satisfied. In this case we have
\begin{eqnarray}
{\tilde X}_i={\tilde x}_i-f\left(\frac{t}{\tau}\right)\frac{\gamma_{ij}\tilde{P}_j}{2M}.{}\label{repch}
\end{eqnarray}
 So, if relation (\ref{cond}) holds the dependence of  coordinates on the relative momenta vanishes and  as was shown in the Section 2 the motion of the center-of-mass is independent of the relative motion.

\section{Conclusion}
Twist-deformed space-time characterized by commutation relations (\ref{ntp})-(\ref{nt1}) has been considered.
  We have studied a general case when coordinates corresponding to different particles satisfy noncommutative algebra  with different parameters (\ref{ntp})-(\ref{nt1}).

  Features of motion of composite system have been examined in the space.
The total momentum defined as sum of momenta of particles in the system is integral of motion in twist-deformed space. Therefore, the coordinates and momenta of the center-of-mass and coordinates and momenta of the relative motion have been introduced in traditional way (\ref{m}), (\ref{m1}), (\ref{06}).
We have shown that commutation relations for coordinates of the center-of-mass of a composite system depend on its mass and composition. In particular case of system made of $N$ particles of the same masses  we have shown that the parameter of noncommutativity corresponding to the motion of the center-of-mass is reduced by the factor $1/N$ in comparison with parameters of noncommutativity corresponding to  particles which form the system (\ref{red}).  So, influence of features of space structures in the Planck scale on composite systems is less than this influence on elementary particles.
 In addition we have found that commutator for coordinates of the center-of-mass and coordinates of the relative motion is not equal to zero (\ref{cr}) because of noncommutativity (\ref{ntp}). So,  the motion of the center-of-mass depends on the relative motion in twist-deformed space.

Also, the motion of a particle (composite system) has been studied in gravitational field in twist-deformed space. We have shown that the weak equivalence principle is not preserved in the frame of commutation relations (\ref{ntp})-(\ref{nt1}). The motion of a particle (composite system) in gravitational field in twist-deformed space depends on its mass and composition.

We have found that if commutator for coordinates of a particle is proportional inversely to its mass (\ref{cond}) one has a) commutator for coordinates of the center-of-mass of a composite system is proportional inversely to its total mass and does not depend on the system's composition; b) motion of the center-of-mass is independent of the relative motion; c) the weak equivalence principle is recovered; d) coordinates  can be considered as kinematic variables in twist-deformed space.

It is worth noting that condition (\ref{cond}) is important for preserving of the weak equivalence principle   in noncommutative space of canonical type \cite{GnatenkoPLA13,GnatenkoMPLA17}, in noncommutative space with rotational symmetry \cite{GnatenkoIJMPA18,GnatenkoEPL18}, in Lie-algebraic noncommutative space \cite{Gnatenko_arxiv18}. Also idea to relate parameters of deformed algebra with mass was proposed in \cite{Tkachuk,Quesne,Tkachuk1} for solving problems in deformed space with minimal length.

\section*{Acknowledgements}

The author thanks Prof. V. M. Tkachuk for his
advices and  support during research.
This work was partly supported by the Project $\Phi\Phi$-63Hp
(No. 0117U007190) from the Ministry of Education
and Science of Ukraine and the grant of the President of Ukraine for support of scientific researches of young scientists (F-75).

\end{document}